\documentclass[aps,prb,reprint,twocolumn,groupedaddress,showpacs]{revtex4-1}
\usepackage{amsmath}
\usepackage{amssymb}
\usepackage{graphicx}
\usepackage{color} %required to colour text
\usepackage[normalem]{ulem} %required for the \sout command to strikeout text
\newcommand{\eps}{\epsilon}
\newcommand{\veps}{\varepsilon}

\newcommand{\etal}{\emph{et al}.}
\newcommand{\nth}[1]{#1^\text{th}}
\newcommand{\ket}[1]{\left|#1\right\rangle}
\newcommand{\bra}[1]{\left\langle #1\right|}
\newcommand{\ketbra}[2]{\left|#1\right\rangle \!\! \left\langle #2\right|}

\renewcommand{\vec}[1]{\mathbf{#1}}

\begin{document}

%\preprint{}

\title{Anomalous magnetotransport through reflection-symmetric artificial molecules}

% repeat the \author . \affiliation etc. as needed

\author{B. D'Anjou}
\author{W. A. Coish}
%\thanks{}
\affiliation{Department of Physics, McGill University, Montreal, Quebec H3A 2T8, Canada}
%\email{benjamin.danjou@mail.mcgill.ca} % I recommend we don't include this since e-mail addresses change and can always easily be found with Google.


\date{\today}

\begin{abstract}
We calculate magnetotransport oscillations in current through a triple-quantum-dot molecule, accounting for higher harmonics (having flux period $h/ne$, with $n$ an integer). For a reflection-symmetric triple quantum dot, we find that harmonics with $n$ odd can dominate over those with $n$ even. This is opposite to the behavior theoretically predicted due to `dark-state' localization, but has been observed in recent experiments [L. Gaudreau \etal, Phys. Rev. B, 80, 075415 (2009)], albeit in a triple-dot that may not exhibit reflection symmetry. This feature arises from a more general result: In the weak-coupling limit, we find that the current is flux-independent for an arbitrary reflection-symmetric Aharonov-Bohm network. We further show that these effects are observable in nanoscale systems even in the presence of typical dephasing sources.

\end{abstract}

\pacs{73.63.Kv, 73.23.-b} %added code for magnetoresistance
%\keywords{}

\maketitle

%\section{Introduction \label{sec:Intro}}
 
\section{Introduction}
The Aharonov-Bohm effect has been the subject of sustained experimental and theoretical investigation since it was first predicted in the middle of the $\nth{20}$ century~\cite{Ehrenberg1949,Aharonov1959}. This effect is most commonly observed in transport through mesoscopic rings threaded by a magnetic flux, $\Phi$, in which the current or conductance is periodically modulated as $\Phi$ is varied~\cite{Pannetier1984,Webb1985,Chandrasekhar1985,Umbach1986,Mankiewich1988}. For free electron waves propagating in both arms of a ring, the periodicity of these oscillations is typically the Dirac flux quantum, $\Phi_0 = h/e$. However, weak localization due to disorder can also give rise to oscillations of period $\Phi_0/2$~\cite{Altshuler1981} known as Altshuler-Aronov-Spivak (AAS) oscillations. The interplay of these two effects has been studied theoretically~\cite{Stone1986} and controlled experimentally by embedding a quantum dot in 
one arm of the interferometer~\cite{Yacoby1996,Hackenbroich1996}. More recently, the fabrication of smaller and cleaner rings has allowed for the observation of higher harmonics---oscillations of period $\Phi_0/n$, with $n>2$~\cite{Hansen2001,Grbic2008}---associated with electrons circling the ring $n$ times. Trajectories that circle the ring multiple times require longer coherence lengths to demonstrate robust interference and are therefore more susceptible to dephasing, typically leading to a decay of the harmonics with increasing $n$ \cite{Hansen2001}. The smallest and simplest ring in which one could expect to observe these higher harmonics consists of a molecule of three sites in a triangular arrangement (provided, e.g., by quantum dots~\cite{Gaudreau2009}, atoms, or implanted donor impurities~\cite{Fuechsle2012}). 

Recent coherent magnetotransport measurements performed on a triple-quantum-dot device suggest, surprisingly, that the harmonics do not decay monotonically with increasing $n$~\cite{Gaudreau2009}. Theory has predicted the formation of a localized dark state in the triple dot, leading to a dominant $n=2$ harmonic~\cite{Emary2007,Delgado2007,Delgado2008,Poltl2009,Weymann2011,Dominguez2011} as in AAS oscillations~\cite{Carini1984}. In contrast, the results of Ref.~\cite{Gaudreau2009} indicate that the $n=3$ harmonic can be dominant over that with $n=2$. To the best of our knowledge, a detailed theory of the $n>2$ harmonics in these systems has not yet been given. Because of their greater sensitivity to electric and magnetic field fluctuations, understanding the large-$n$ harmonics in small molecular rings may allow for enhanced magnetic-field and noise sensing in future nanoscale devices, similar to recent proposals for mesoscopic systems \cite{Strambini2010}. Moreover, in contrast with the case of mesoscopic rings, we show that discrete symmetries associated with finite-dimensional molecular systems can lead to non-trivial features in magnetotransport.
\begin{figure}%removed figure placement commands -- better to let RevTeX decide
\centering
\includegraphics[width=\columnwidth]{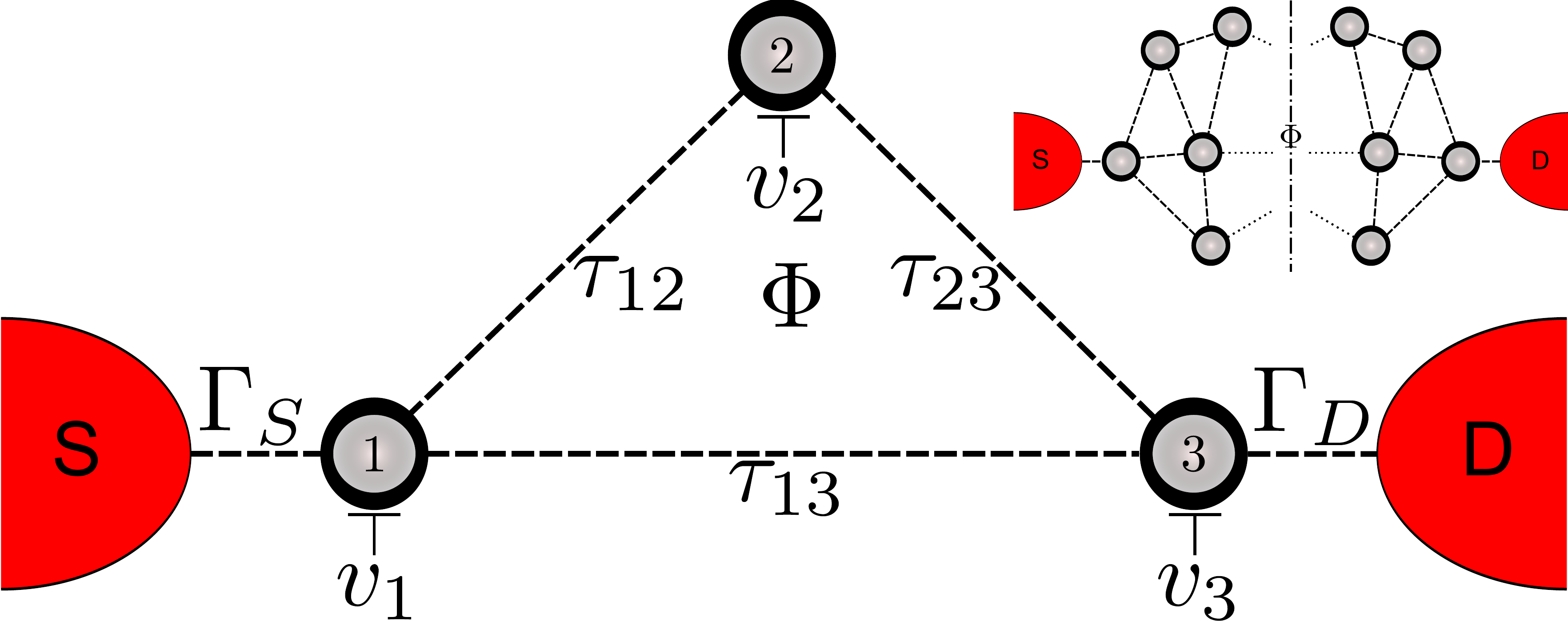}
\centering
\caption{(Color online) Three quantum dots in a ring arrangement, with dots $1$ and $3$ connected to the source $S$ and drain $D$ via tunneling rates $\Gamma_S$ and $\Gamma_D$, respectively. Dot $i$ is subject to a local potential $v_i$ (taken here to be $v_i = 0$). The lowest single-particle orbitals of dots $i$ and $j$ are connected via tunnel couplings $\tau_{ij}$. The ring is threaded by a flux $\Phi$. Inset: A reflection-symmetric Aharonov-Bohm network. \label{fig:TripleDot}}
\end{figure}

For definiteness, we focus on the triple-dot arrangement shown in Fig. \ref{fig:TripleDot}, although several results extend naturally to a larger number of sites and more general geometries. An accurate description of the large-$n$ harmonics necessarily requires a theory that accounts for non-linear response (large voltage bias), that goes beyond the leading order in weak-coupling to leads, and that accounts for strong Coulomb interactions. A coherent master equation similar to that employed in Ref. \cite{Emary2007} can be rigorously justified in an experimentally accessible regime and satisfies all the above criteria. We show that for a reflection-symmetric configuration of the triple dot, the symmetry of the eigenstates forbids localization that would lead to $n=2$ AAS-like oscillations. This effect arises from a more fundamental result: In the weak-coupling limit ($\Gamma/\Delta\varepsilon\to 0$ with tunneling rate $\Gamma$ and molecular level spacing $\Delta\varepsilon$), we find that the current through an \emph{arbitrary} reflection-symmetric Aharonov-Bohm network (inset of Fig.~\ref{fig:TripleDot}) is magnetic-field independent. 

\section{Transport model}
In Fig. \ref{fig:TripleDot}, we consider large single-dot level spacing, $\omega_0>\Delta\mu,\,U,\,U'$, with $\Delta\mu=\mu_S-\mu_D$ the bias for source(drain) at chemical potential $\mu_{S(D)}$ and $U,\,U'$ on-site and nearest-neighbor charging energies, respectively (setting $e=\hbar=1$). For simplicity, we take all dots to be at the same potential, $v_i=0$, and choose $\mu_S>0,\,\mu_D<0$. In this regime only the lowest-energy dot orbitals ($i=1,2,3$) participate. Orbitals $i$ and $j$ are connected via a tunnel coupling $\tau_{ij}$, and dots $1$ and $3$ are coupled to the source and drain, respectively.

We choose the gauge such that $\tau_{12}$ and $\tau_{23}$ are real, while $\tau_{13}=|\tau_{13}| e^{-2\pi i \phi}$, with $\phi = \Phi/\Phi_0$ the reduced flux. The full Hamiltonian is
\begin{align}\label{eq:Hamiltonian}
 \mathcal{H} &= \mathcal{H}_\mathrm{3D} + \mathcal{H}_L + \mathcal{H}_{3DL},\\
 \mathcal{H}_{3D} &= \sum_{i\ne j,\sigma}\tau_{ij}d_{i\sigma}^\dagger d_{j\sigma}+\frac{\epsilon_z}{2}\sum_i (n_{i\uparrow}-n_{i\downarrow})+ \mathcal{H}_C,\\
 \mathcal{H}_L =& \sum_{lk\sigma} \epsilon_{lk} c_{lk\sigma}^\dagger c_{lk\sigma};
 \mathcal{H}_{3DL} = \sum_{ilk\sigma} t_{ilk}d_{i\sigma}^\dagger c_{lk\sigma}+\mathrm{h.c.}.
\end{align}
Here, $\epsilon_z$ is the Zeeman splitting, the Coulomb interaction is $\mathcal{H}_C=U\! \sum_{i} n_{i\uparrow} n_{i\downarrow} + U' \! \sum_{i > j, \sigma\sigma'} \! n_{i\sigma} n_{j\sigma'}$, $d_{i\sigma}$ annihilates an electron on dot $i$ with spin $\sigma$, and $n_{i\sigma} = d_{i\sigma}^\dag d_{i\sigma}$. The operator $c_{lk\sigma}$ annihilates an electron in state $k$ of energy $\eps_{lk}$ and spin $\sigma$ in lead $l=S,D$. We neglect diamagnetic effects on $|\tau_{ij}|$, but account for the Peierls phase. 

We consider the Coulomb-blockade regime, $\mu_S<U,U'$, in which energy-conserving transitions involve the vacuum and one-electron subspace. Restricting to one electron, the three localized orbitals hybridize into molecular orbitals with eigenenergies $\varepsilon_{k\sigma}$ for $k=1,2,3$. We focus on the experimentally relevant high-bias regime, $\mu_S-\varepsilon_{k\sigma}>k_\mathrm{B} T$, $\varepsilon_{k\sigma}-\mu_D>k_\mathrm{B} T$, where all one-electron states are accessible via sequential-tunneling processes with leads at temperature $T$. We further take the density of states, $\nu_{kl}$, and dot-lead tunnel couplings, $t_{ilk}$, to be spin- and energy-independent over the bias $\Delta\mu$: $\nu_{lk}\approx\nu_l$ and $t_{ilk}\approx t_{il}$, leading to tunneling rates $\Gamma_S = 2\pi \nu_S |t_{1S}|^2$ and $\Gamma_D = 2\pi \nu_D |t_{3D}|^2$. 

Dynamics giving rise to magnetotransport oscillations in this system are highly sensitive to sources of dephasing, so it is essential to establish an accurate regime of validity for the associated equation of motion, particularly in the chosen high-bias regime. To approach the problem systematically, we start from the exact Nakajima-Zwanzig integro-differential equation~\cite{Breuer2002} for the full dot-lead density matrix $\varrho(t)$~\cite{Vaz2010}. From this, we find an equation for $\rho$, an effective reduced triple-dot density matrix in terms of spinless fermions. The dynamics of $\rho$ are controlled by dot-lead correlation functions, which we evaluate within a Born approximation (leading order in $\mathcal{H}_{3DL}$). This approximation is valid even for strong coupling, $\Gamma_l \gg |\tau_{ij}|$, as long as higher-order cotunneling processes can be neglected (see below). In the high-bias regime considered here, the dot-lead correlation time, $\tau_c\sim 1/|\mu_l-\varepsilon_{k\sigma}|$, is taken to be much shorter than the characteristic evolution time of $\rho$ [$\gtrsim \mathrm{min}\left(1/|\tau_{ij}|,1/\Gamma_l\right)$]. In this singular-coupling limit~\cite{Spohn1980,Schultz2009}, a Markov approximation is justified. We thus obtain a coherent master equation:
\begin{equation}
	\dot{\rho} = -i \left[H,\rho\right] + 2\Gamma_S \mathcal{D}[d_1^\dagger]\rho + \Gamma_D \mathcal{D}[d_3]\rho.\label{eq:MasterEquation}
\end{equation}
Here, $H=\sum_{i\ne j}\tau_{ij} d_i^\dagger d_j$ is the triple-dot Hamiltonian, where $d_i^\dagger$ creates a spinless fermion on dot $i$, $d_i^\dagger\left|0\right>$=$\left|i\right>$. The superoperator defined by $\mathcal{D}[\mathcal{O}]\rho = \mathcal{O}\rho\mathcal{O}^\dag - \frac{1}{2}\left\{\mathcal{O}^\dag\mathcal{O},\rho\right\}$ is the Lindblad dissipator~\cite{Breuer2002}. The first term in Eq. \eqref{eq:MasterEquation} describes free evolution on the triple dot, while the two following terms describe tunneling into and out of the triple dot. The factor $2$ in the second term accounts for the two possible spin states into which an electron can tunnel. Contributions at higher order in $\mathcal{H}_{3DL}$ can lead to dephasing and suppress the coherent effects described by Eq.~\eqref{eq:MasterEquation}, especially in the considered high-bias regime. However, when $\Gamma_D\lesssim \mathrm{min}\left(|\tau_{ij}|,\Gamma_S\right)$, we find that these cotunneling processes give a small correction in the considered regime, provided $\Gamma_{D} > \mathrm{max}\left\{\left(\Gamma_S/\Delta\mu\right)^2|\tau_{ij}|, \left[\Gamma_S\Gamma_D/(U')^2\right]\Delta\mu\right\}$.

 The current is $I=\text{Tr}\left\{\mathcal{I}\bar{\varrho}\right\}$, where $\bar{\varrho}=\lim_{\tau_m\to\infty}(1/\tau_m)\int_0^{\tau_m} dt \varrho(t)$ is the full stationary dot-lead density matrix with averaging (measurement) time $\tau_m$. Here, the current operator is $\mathcal{I}=\dot{\mathcal{N}}_D=i\left[\mathcal{H},\mathcal{N}_D\right]$, where $\mathcal{N}_D=\sum_{k\sigma}c_{Dk\sigma}^\dagger c_{Dk\sigma}$. When cotunneling corrections are negligible, we find that $I(\phi)$ is given directly from the stationary population $\bar{\rho}_{33}$ of dot $3$, 
\begin{align}
	I(\phi) = \Gamma_D \bar{\rho}_{33}(\phi),\quad \hat{I}_n = \int_{0}^{1} d\phi\, I(\phi) e^{2\pi i n \phi}. \label{eq:Current}
\end{align}

\section{Magnetocurrent harmonics}
\begin{figure}%removed figure placement commands -- better to let RevTeX decide
\centering
\includegraphics[width=\columnwidth]{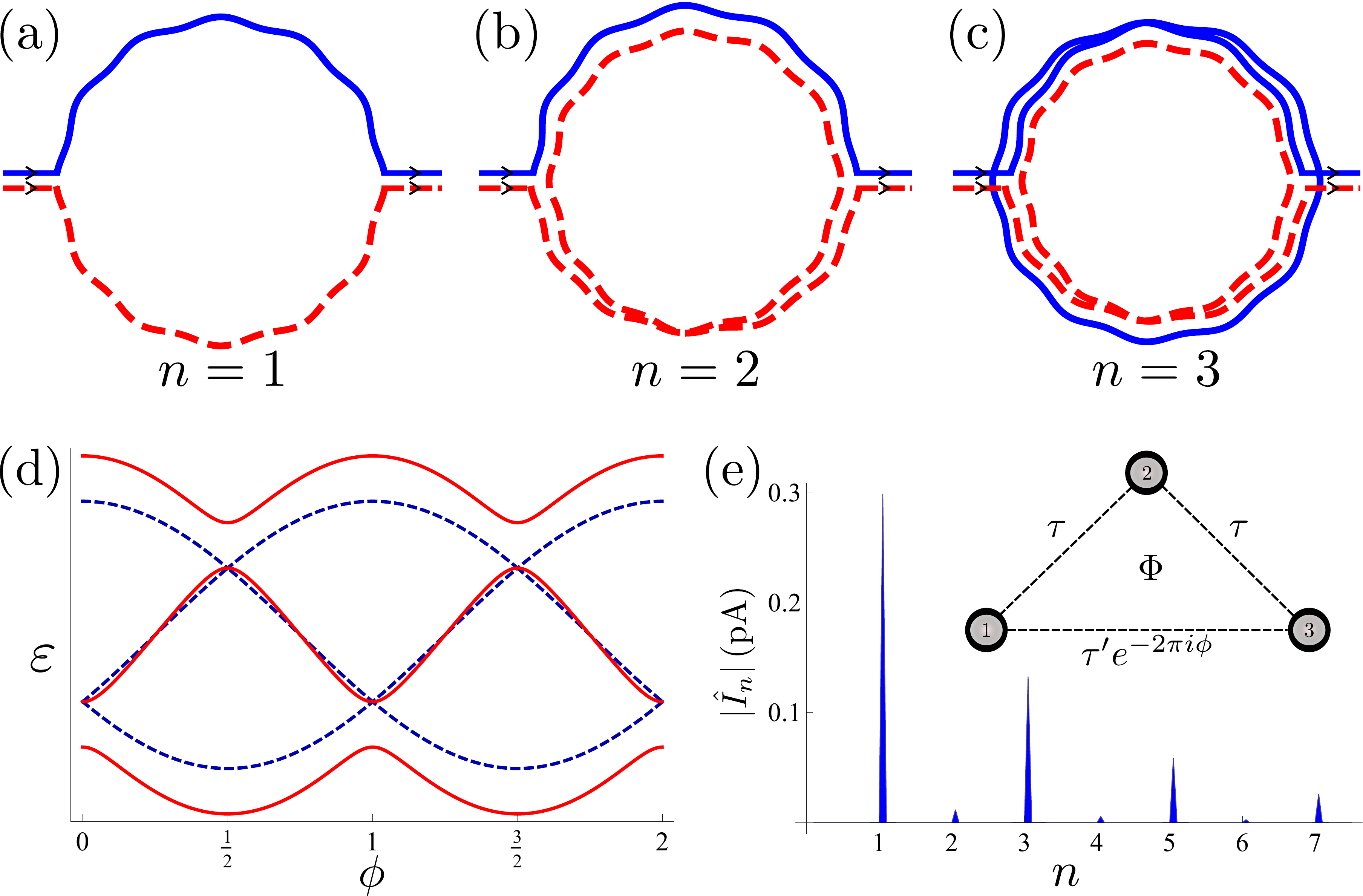}
\centering
\caption{(Color online) (a)-(c) Trajectories leading to the $n=1,2,3$ harmonics. For $n$ even, the transit times for the two paths are always different, suppressing the associated interference. (d) Eigenenergies $\veps$ of the triple dot as a function of the reduced flux $\phi$ for $\tau=\tau_{12}=\tau_{23}=|\tau_{13}|$ (dashed) and for $\tau=\tau_{12}=\tau_{23}\neq|\tau_{13}|=\tau'$ (solid). A gap opens at values of $\phi=m/2$ with $m\in\mathbb{Z}$, protecting the state from localization. To account for the positive effective mass of the electron in the conduction band, the couplings $\tau_{ij}$ should be chosen negative; however, since this does not affect the harmonics of Fig.~\ref{fig:SpectrumAndPaths}(e), we chose them positive for convenience. Choosing $\tau_{ij}<0$ would, however, shift the energy spectrum relative to that shown in (d) by half a flux quantum (as in, e.g., Fig.~1 of Ref.~\cite{Carini1984}). (e) Magnitude $|\hat{I}_n|$ for $\tau=30\,\mu\mathrm{eV}$, $\tau'=20\,\mu\mathrm{eV}$, $2\Gamma_S=20\,\mu\mathrm{eV}$, and $\Gamma_D=1\,\mu\mathrm{eV}$. The $n$-even harmonics are strongly suppressed with respect to those with $n$ odd. \label{fig:SpectrumAndPaths}}
\end{figure}

The behavior of the harmonics, $\hat{I}_n$, will generally depend on the choice of $\tau_{ij}$. When all $\tau_{ij}$ are equal, and for most generic choices of $\tau_{ij}$, the $n$-even harmonics, $\hat{I}_{2k}$, describing AAS-like oscillations, typically dominate in the absence of dephasing \cite{Emary2007,Delgado2007,Delgado2008,Poltl2009,Weymann2011,Dominguez2011,Carini1984}. This behavior contrasts with that suggested by recent experiments~\cite{Gaudreau2009}. We have found, however, that the $n$-even harmonics can be suppressed for a range of tunnel couplings satisfying $|\tau_{12}|\simeq |\tau_{23}| \neq |\tau_{13}|$~\footnote{This differs from the situation reported in the experiments of Ref.~\cite{Gaudreau2009} which motivated this work, where the authors concluded that $|\tau_{13}|\simeq\tau_{23}\neq\tau_{12}$ instead. Furthermore, the authors of Ref.~\cite{Gaudreau2009} report that both dots 1 and 2 were connected to the source. As long as $\tau=\tau_{12}=\tau_{23}\neq \tau'=|\tau_{13}|$, we find that for $x=\tau/\tau' \sim 1$, $|\hat{I}_3|>|\hat{I}_2|$ as long as $\Gamma_S^2/\Gamma_S^1 \ll \Gamma_D/\tau'$, where $\Gamma_S^i$ is the tunneling rate into dot $i$.}. As we will show, this suppression of AAS-like oscillations is a feature unique to low-symmetry molecular systems and distinguishes this case from mesoscopic rings with a near-continuum of orbital states. To see this more explicitly, from Eqs. \eqref{eq:MasterEquation} and \eqref{eq:Current} we find a simple analytical expression for $I$ when $\tau_{12}=\tau_{23}=\tau,\,|\tau_{13}|=\tau'$:
\begin{align}
I(\phi) &= \Gamma_D\left\{3+ z + \frac{\left[1+x^2\right]\left[y\sin(2\pi\phi)+y^2/2\right]}{1+x^4-2x^2\cos(4\pi\phi)} \right\}^{-1}, \nonumber\\
x &= \tau/\tau',\quad y = \Gamma_D/\tau',\quad z = \Gamma_D/2\Gamma_S.\label{eq:ExplicitCurrent}
\end{align}
The harmonics, $\hat{I}_n$, resulting from Eq. \eqref{eq:ExplicitCurrent}, are shown in Fig.~\ref{fig:SpectrumAndPaths}(e) for typical experimentally realizable parameters, displaying a non-monotonic dependence on $n$ (in particular, $|\hat{I}_3|>|\hat{I}_2|$), as in the experimental findings of Ref.~\cite{Gaudreau2009}. Note that Eq.~\eqref{eq:ExplicitCurrent} does not satisfy the Onsager relation, i.e., $I(\phi)\ne I(-\phi)$, but the Onsager relation is generally not obeyed in the nonlinear (high-bias) regime considered here~\cite{Bruder1996}.

The harmonics $\hat{I}_n$ arise from paths that combine to encircle the ring $n$ times [see Figs. \ref{fig:SpectrumAndPaths}(a)-(c)]. Trajectories circling the ring multiple times become more significant when the dwell time on the ring $\sim 1/\Gamma_D$ is large compared to the timescale for coherent evolution $1/\Delta\varepsilon\sim 1/\tau'$, so it is useful to consider an expansion in $y = \Gamma_D/\tau'$. The leading term, $I\simeq \Gamma_D/(3+z)\propto y^0$, corresponds to the current from an incoherent (Pauli) master equation. The first subleading term $(\propto y)$ describes harmonics with $n$ odd and the next order $(\propto y^2)$ harmonics with $n$ even. When $\Gamma_D/\tau'=y\ll 1$, the $n$-odd contributions dominate. The description of these higher harmonics requires that we go beyond the weak-coupling limit, i.e., beyond $\mathcal{O}(y^0)$~\cite{Hackenbroich1996}.

We note that for $x=\tau/\tau'\to 0$ (resulting in 1D transport through dots 1 and 3) the current, Eq.~\eqref{eq:ExplicitCurrent}, still depends on $\phi$ although there is only one path in this case, thus no flux enclosed. This seemingly unphysical result is a consequence of the non-commuting limits $x\to 0$ and $\tau_m\to\infty$. We find that for any finite measurement time $\tau_m$, $I$ is $\phi$-independent when $x\to 0$. Similarly, for any vanishingly small but finite $x$, the current $I$ will have the $\phi$-dependence indicated in Eq.~\eqref{eq:ExplicitCurrent} in the limit $\tau_m\to\infty$. 

In general, we find it convenient to describe magnetotransport oscillations in terms of two distinct contributions. The first, \emph{static} contribution, gives rise to AAS-like oscillations due to localization and delocalization of molecular eigenstates as the magnetic field is varied. These static contributions can be found accurately from an incoherent (Pauli) master equation, which neglects off-diagonal elements of $\rho$ in the eigenbasis of $H$ and gives magnetotransport oscillations at leading order in a conventional weak-coupling expansion, $\mathcal{O}(y^0)$. The second, \emph{dynamical}, contributions give rise to the standard Aharonov-Bohm effect and its harmonics associated with the coherent motion of electrons circling the ring. The dynamical contributions arise only at subleading order in a weak-coupling expansion [$\sim\mathcal{O}(y)$ or higher]. 

We can understand the unusual behavior shown in Fig.~\ref{fig:SpectrumAndPaths}(e) by accounting for both the static and dynamical contributions. The static contributions are best analyzed within degenerate perturbation theory~\cite{Carini1984}. When the couplings have equal magnitude, $|\tau_{ij}|=\tau\;\forall i,j$, the eigenstates are completely delocalized molecular states, pairs of which are degenerate for $\phi=m/2$, $m\in \mathbb{Z}$ [see Fig.~\ref{fig:SpectrumAndPaths}(d)]~\cite{Delgado2008}. Introducing a small real perturbation to the triple-dot Hamiltonian taking it away from $|\tau_{ij}|=\tau\;\forall i,j$, leads to eigenstates that are symmetric and antisymmetric linear combinations of the degenerate states at leading order in degenerate perturbation theory. These combinations are typically localized, strongly suppressing current at near-degeneracies, when $\phi=m/2$, and hence inducing AAS-like oscillations of period $\Phi_0/2$~\cite{Carini1984,Emary2007,Delgado2007,Delgado2008,Poltl2009,Weymann2011}. However, when $\tau\equiv\tau_{12}=\tau_{23}$ and $\tau\neq\tau'\equiv|\tau_{13}|$, i.e. when $\left[H,\Theta\right]=0$, $\Theta=\Pi K$ being an anti-unitary operator composed of the parity operator $\Pi = \ketbra{1}{3}+\ketbra{3}{1}+\ketbra{2}{2}$ and of the complex-conjugation operator $K$, we exploit this special discrete symmetry of the triple dot \cite{Oguri2011} to gain further insight. In this case the eigenstates $\ket{\veps_p}= \sum_i c_p^i \ket{i}$ of $H$ are simultaneous eigenstates of $\Theta$, uniformly delocalized across dots $1$ and $3$ ($|c_p^1|=|c_p^3|\,\forall\,p$), strongly suppressing the AAS-like oscillations. 
Indeed, a state $\ket{\veps_p}$ is loaded at a rate $2\gamma_S^p = 2|c_p^{1}|^2 \Gamma_S$ and unloaded at a rate $\gamma_D^p = |c_p^3|^2 \Gamma_D$. Solving the steady-state Pauli master equation $2\gamma_S^p \rho_{00} - \gamma_D^p \rho_{pp}=0$ to obtain the `static' contribution and using $|c_p^3|^2=|c_p^1|^2$, we find that $\rho_{pp}/\rho_{00}=2\Gamma_S/\Gamma_D$ is the same and independent of $\phi$ $\forall\,p$. Because $\sum_p\rho_{pp}=1$, this means that the current, $I=\Gamma_D \bar{\rho}_{33}$, is also independent of $\phi$. We emphasize that this $\phi$-independence in the weak-coupling limit, $\Gamma_D/\Delta\varepsilon\to 0$, holds for \emph{any} Aharonov-Bohm network with a single dot connected to each lead whenever the dot configuration has mirror symmetry (inset of Fig. \ref{fig:TripleDot}). Thus, when $\left[H,\Theta\right]=0$, any Aharonov-Bohm oscillations must arise from coherent dynamics on the triple dot contained in the first term on the right-hand side of Eq.~\eqref{eq:MasterEquation} and not from the localization/delocalization of eigenstates responsible for AAS-like oscillations. If a perturbation $V$ breaks parity at $\phi=m/2$ ($\left[V,\Pi\right]\neq 0$), lowest-order perturbation theory shows that the eigenstates will remain delocalized as long as $|\bra{\veps_+}\!V\!\ket{\veps_-}|/\Delta\ll 1$, where $\ket{\veps_+}$ and $\ket{\veps_-}$ are the nearly degenerate eigenstates of $H$ and $\Pi$ at $\phi=m/2$ and where $\Delta=\left|3|\tau'|-\sqrt{|\tau'|^2+8\tau^2}\right|/2$ is the gap opened at the degeneracy point [see Fig.~\ref{fig:SpectrumAndPaths}(d)]. 
Even when the AAS-like oscillations are suppressed, we might still expect to find a significant dynamical contribution to the $n=2$ harmonics coming from the interference of paths circling the ring twice [see Fig.~\ref{fig:SpectrumAndPaths}(b)]. However, because backscattering is forbidden in the high-bias regime, the remaining asymmetric paths result in different transit times, reducing interference. In contrast, the $n$-odd contributions can arise from paths that are symmetric in both arms [see Figs.~\ref{fig:SpectrumAndPaths}(a) and (c)]. 

\section{Effect of dephasing}
\begin{figure}%removed figure placement commands -- better to let RevTeX decide
\centering
\includegraphics[width=\columnwidth]{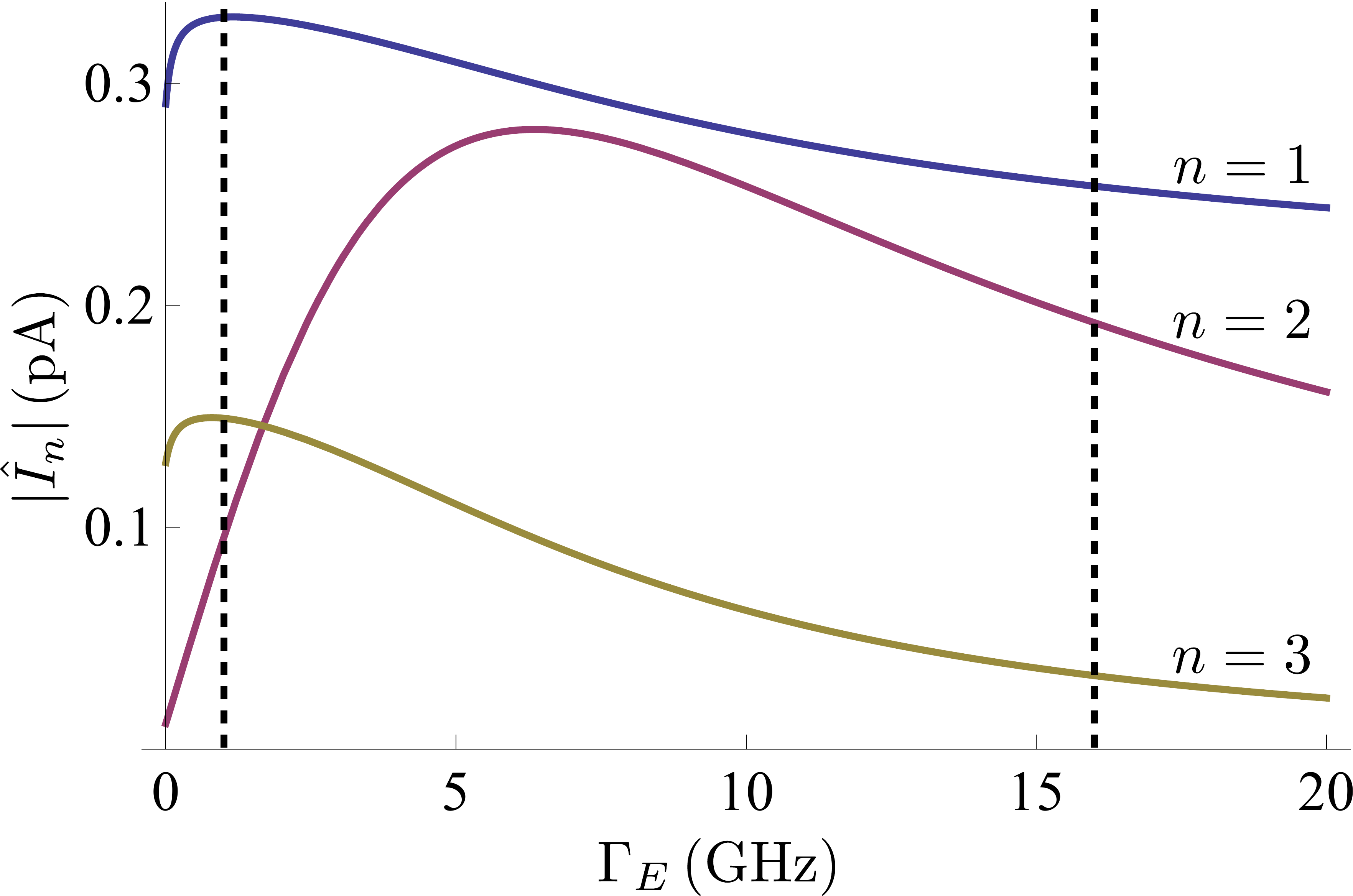}
\centering
\caption{(Color online) Magnitude $|\hat{I}_n|$ of the $n=1,2,3$ harmonics as a function of the dephasing rate, $\Gamma_E$, for $\tau_{12}=\tau_{23}=30\,\mu\mathrm{eV}$, $|\tau_{13}|=20\,\mu\mathrm{eV}$, $2\Gamma_S=20\,\mu\mathrm{eV}$ and $\Gamma_D=1\,\mu\mathrm{eV}$. When $\Gamma_E\gg|\tau_{ij}|$, $|\hat{I}_n|$ decreases monotonically with $n$. The first vertical line (at $\Gamma_E=1\,\mathrm{GHz}$), corresponds to the measured dephasing rate for a nanoscale double quantum dot of size $\sim250\,\mathrm{nm}$~\cite{Hayashi2003}. The second vertical line (at $\Gamma_E=16\,\mathrm{GHz}$) corresponds to the estimated dephasing time for a mesoscopic ring of size $\sim 1\,\mu\mathrm{m}$~\cite{Hansen2001} and is obtained from our theory by scaling according to the relative dipole moments [$16\,\mathrm{GHz}= (1\,\mu\mathrm{m}/250\,\mathrm{nm})^2 \times 1\,\mathrm{GHz}$]. \label{fig:HarmonicsDephasing}}
\end{figure}

To account for dephasing we add a term $V_E(t) = -e\sum_{i\sigma} \mathbf{E}(t)\cdot\mathbf{r}_i n_{i\sigma}$ to $\mathcal{H}$, Eq.~\eqref{eq:Hamiltonian}. This term describes electric-dipole coupling of localized dot orbitals at positions $\mathbf{r}_i$ to a fluctuating electric field, $\mathbf{E}(t)$, taken to be uniform across the triple dot. The effect of $V_E(t)$ is to introduce a dissipator to Eq.~\eqref{eq:MasterEquation}:
\begin{align}
	\dot{\rho} = \mathcal{L}_0\rho + \sum_{i> j} \Gamma^E_{ij} \mathcal{D}[d_i^\dagger d_i - d_j^\dagger d_j]\rho, \label{eq:MasterEquationDephasing}
\end{align}
where $\mathcal{L}_0$ generates the right-hand side of Eq.~\eqref{eq:MasterEquation}. When $V_E(t)$ dominates over $\mathcal{H}$ and when the noise is Gaussian and white, we find $\Gamma^E_{ij}\sim|\vec{r}_i-\vec{r}_j|^2$, implying that electric-field-induced dephasing is more significant for larger systems. Note that Eq.~\eqref{eq:MasterEquationDephasing} is only strictly valid when $\Gamma^E_{ij}=0$ or $\Gamma^E_{ij}\gg \max(\tau_{ij},\Gamma_l)$. When the dephasing term in Eq.~\eqref{eq:MasterEquationDephasing} is non-zero, it acts as a which-path measurement for both arms of the ring and destroys the interference that would lead to higher harmonics. Fig.~\ref{fig:HarmonicsDephasing} shows $|\hat{I}_n|$ for the simple case where $\Gamma^E_{ij}\equiv\Gamma_E$ for all $i,j$.
Increased dephasing re-establishes the monotonic decrease of $|\hat{I}_n|$ with $n$ when $\Gamma_E \gg |\tau_{ij}|$, i.e., when the characteristic evolution time $|\tau_{ij}|^{-1}$ on the triple dot is longer than the dephasing time ${\Gamma_E}^{-1}$. Thus, we do not expect the non-monotonic behavior of $|\hat{I}_n|$ displayed in Fig.~\ref{fig:SpectrumAndPaths}(a) to be observed in larger rings where the relative electric dipole moment $\sim |\vec{r}_i-\vec{r}_j|$ between arms is much larger.

\section{Conclusion}
In summary, we have derived a non-linear (high-bias) transport theory for a triple quantum dot in the strongly-interacting Coulomb-blockade regime, and have used it to describe $\Phi_0/n$ harmonics of Aharonov-Bohm oscillations beyond the limit of weak coupling to the leads. We have shown that for a reflection-symmetric Aharonov-Bohm network, the current is flux-independent in the weak-coupling limit ($\Gamma/\Delta\varepsilon\to 0$). In the simplest case of a reflection-symmetric triple dot, this results in a strong suppression of the $n$-even harmonics compared to those with $n$ odd. This happens for two reasons: first, the $n$-even AAS-like harmonics due to the localization of triple-dot eigenstates are suppressed since symmetry requires that the eigenstates be delocalized across the triple dot; second, the dynamical paths leading to $n$-even harmonics have different transit times, suppressing interference. These results may help explain experiments such as those of Ref.~\cite{Gaudreau2009}. We have also shown that strong dephasing restores a monotonic suppression of harmonics with increasing $n$. Finally, we remark that these results are important for the design of noise and magnetic-field sensors using Aharonov-Bohm interference~\cite{Strambini2010}. The coherent effects presented here are likely to be substantially stronger in still smaller devices, such as those very recently constructed from individual atoms or donor impurities~\cite{Fuechsle2012}.

\section{Acknowledgments}
We thank A.~Sachrajda and A.~Clerk for useful discussions and acknowledge financial support from NSERC, CIFAR, FQRNT, and INTRIQ.

\bibliography{prl.00}

\end{document}